\documentclass[conference,a4paper]{IEEEtran}
\usepackage{cite}
\usepackage{amsmath}
\usepackage{algorithmic}
\usepackage{array}
\usepackage{mdwmath}
\usepackage{mdwtab}
\usepackage{eqparbox}
\usepackage[tight,footnotesize]{subfigure}
\usepackage[font=footnotesize]{subfig}
\usepackage{fixltx2e}
\usepackage{stfloats}
\usepackage[pdftex]{graphicx}
\usepackage{amssymb}
\begin{document}

\sloppy

\title{On the Cooperative Communication over\\Cognitive Interference Channel}

\author{
  \IEEEauthorblockN{Mohammad Kazemi}

  \IEEEauthorblockA{University of Rochester\\Rochester, NY, USA\\\texttt{mkazemi@ece.rochester.edu}}

 \and

 \IEEEauthorblockN{Sahar Hashemgeloogerdi}

 \IEEEauthorblockA{University of Rochester\\Rochester, NY, USA\\\texttt{shashemg@ece.rochester.edu}}
}

\maketitle

\begin{abstract}
In this paper, we investigate the problem of communication over cognitive interference channel (CIC) with partially cooperating (PC) destinations (CIC-PC). This channel consists of two source nodes communicating two independent messages to their corresponding destination nodes. One of the sources, referred to as the cognitive source, has a noncausal knowledge of the message of the other source, referred to as the primary source. Each destination is assumed to decode only its intended message.
%Source 1, referred to as the cognitive source, knows both messages, whereas source 2, referred to as the primary source, knows only message 2.
In addition, the destination corresponding to the cognitive source assists the other destination by transmitting cooperative information through a relay link. We derive a new upper bound on the capacity region of discrete memoryless CIC-PC. Moreover, we characterize the capacity region for two new classes of this channel: (1) degraded CIC-PC, and (2) a class of semideterministic CIC-PC.%, where the channel output observed by the relay is a deterministic function of the channel inputs.
\end{abstract}

\section{Introduction}
Discrete memoryless cognitive interference channel (CIC) is the channel model introduced in \cite{c1} to investigate the information theoretic limits of communication over cognitive radio networks. Interference, which undeniably arises in cognitive radio networks, affects adversely the data communication rates. Therefore, a model of CIC with partially cooperating (PC) destinations was introduced in \cite{c2}, where relay links are exploited into standard CIC to improve the throughput and reliability through cooperative relaying of information.

In this paper, we study the discrete memoryless CIC-PC which, as shown in Fig. 1, is a network with two sources communicating two independent and uniformly distributed messages to two destinations. Source 1, referred to as the cognitive source, knows both messages 1 and 2, whereas source 2, referred to as the primary source, knows only message 2. \emph{Each destination needs to decode only its intended message}. In addition, destination 1 acts as a standard relay node \cite{c3,c4,c5} and assists destination 2 by transmitting cooperative information through a relay link.

We derive a new upper bound on the capacity region of the discrete memoryless CIC-PC. We then study two classes of this channel. The first class is the degraded CIC-PC, where the channel output at destination 2 is degraded with respect to the channel output at destination 1 (relay). We derive the capacity region for the degraded CIC-PC. The second class is the semideterministic CIC-PC, where the channel output observed by destination 1 (relay) is a deterministic function of the channel inputs. We characterize the capacity region for a variation of this channel where the channel model satisfies a certain condition, ensuring that the destination 1 (relay) can decode better than destination 2. We refer to this condition as the "\emph{more capable}" regime. We show that other classes of semideterministic CIC-PC for which the capacity region had been established are special cases of more capable semideterministic CIC-PC.

A different model of CIC-PC was also studied in \cite{c6,c7,c8}, in which the message sent by the primary source is decoded by both decoders. Therefore, in this channel model, the primary source node is not a source of interference on the communication of cognitive source and its corresponding destination.

The rest of the paper is organized as follows. In Section II, we provide a formal definition for the discrete memoryless CIC-PC. In Section III, we establish a new upper bound on the capacity region of this channel. Finally, in Section IV, we characterize the capacity region of the degraded CIC-PC and more capable semideterministic CIC-PC.

%In this paper, we study the discrete memoryless CIC-PC. channel model introduced in \cite{c1}.
%The discrete memoryless cognitive interference channel (CIC) was the channel setup introduced in \cite{c1} to study the fundamental limits of communication over cognitive radio networs. Interference, which undeniably arises in cognitive radio networks, affects adversely the data communication rates. Therefore, cooperative relay CIC (RCIC) was introduced in \cite{c3,c4} to study cooperative relaying of information as a powerful technique to improve the reliability and throughput of CICs.

\begin{figure}\label{fig:test}
  \centering
  \includegraphics[scale=.24]{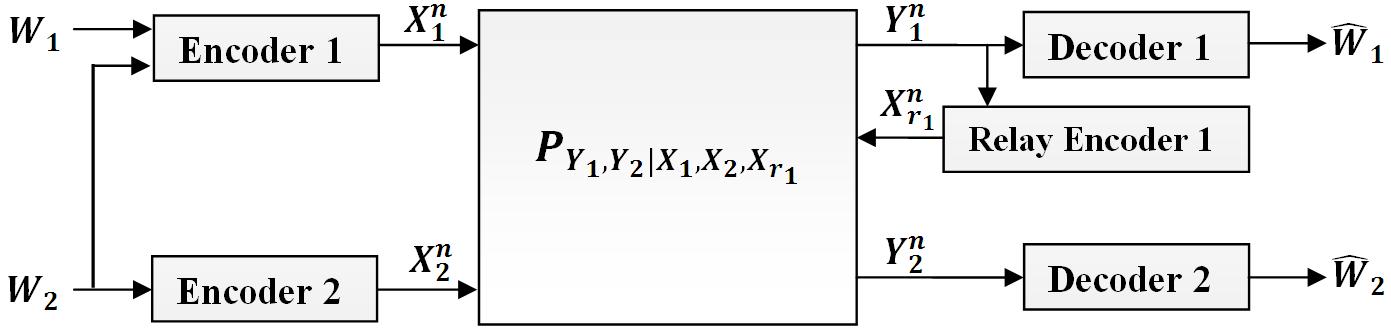}
  \caption {Discrete memoryless CIC with PC destinations (CIC-PC)}
  \label{figurelabel}
\end{figure}

\section{Notations and Definitions}
\subsection{Notations}
Random variables (RVs) are indicated by upper case letters, (e.g. $X$) and their realizations are shown by the respective lower case letters (e.g. $x$). The probability mass function (p.m.f.) of a random variable $X$ over its corresponding finite alphabet set $\mathcal{X}$ is indicated by $p_{X}(x)$ where occasionally subscript $X$ is omitted. The conditional p.m.f. of a random variable $X$ given random variable $Y$ is denoted by $p_{X|Y}(x|y)$. Random vector $(X_1,X_2,...,X_n)$ is indicated by $X^n$ or \emph{\textbf{X}}; a sequence of random variables $(X_i,X_{i+1},...,X_{j-1},X_j)$ is denoted by $X_{i}^{j}$. For brevity, $X^j$ is used instead of $X_{1}^{j}$. Entropy of a RV and mutual information between two RVs are indicated by $H(\cdot)$ and $I(\cdot;\cdot)$, respectively.

\subsection{Definitions}
\textbf{Definition 1.} Discrete memoryless CIC-PC, as shown in Fig. 1, consists of two finite discrete source input alphabets $\mathcal{X}_1$, $\mathcal{X}_2$, a finite discrete relay input alphabet $\mathcal{X}_{r_1}$, two discrete output alphabets $\mathcal{Y}_1$ and $\mathcal{Y}_2$ and a set of transition probability distributions $p(y_1,y_2|x_1,x_2,x_{r_1})$ describing the relationship between transmitted symbols $(x_1,x_2,x_{r_1})\in{\mathcal{X}_1\times{\mathcal{X}_2}\times{\mathcal{X}_{r_1}}}$ and received symbols $(y_1,y_2)\in{\mathcal{Y}_1\times{\mathcal{Y}_2}}$.\vspace{1.5mm}

\textbf{Definition 2.} A $(2^{nR_1},2^{nR_2},n)$ code for the CIC-PC consists of the following: (1) Two message sets $\mathcal{W}_i=\{1,2,...,2^{nR_i}\}$, $i = 1, 2$; (2) Two messages $W_1$ and $W_2$ which are independent random variables uniformly distributed over $\mathcal{W}_1$ and $\mathcal{W}_2$, respectively; (3) encoder $f_1:\mathcal{W}_1\times{\mathcal{W}_2}\rightarrow{\mathcal{X}_{1}^n}$, which maps message pair $(\omega_1,\omega_2)\in{\mathcal{W}_1\times{\mathcal{W}_2}}$ to a codeword $x_1^n\in{\mathcal{X}_1^n}$, encoder $f_2:\mathcal{W}_2\rightarrow{\mathcal{X}_2^n}$ which maps message $\omega_2\in{\mathcal{W}_2}$ to a codeword $x_2^n\in{\mathcal{X}_2^n}$ and a set of relay functions $\{\varphi_i\}_{i=1}^n$ such that $x_{r_1,i}=\varphi_i(y_{1,1}^{i-1})$, $i=1,...,n$; (4) Decoder $g_i:\mathcal{Y}_i^n\rightarrow{\mathcal{W}_i}$, which maps a received sequence $y_i^n$ to a message $\widehat{\omega}_i\in{\mathcal{W}_i}$, $i = 1, 2$.\vspace{1.5mm}

\textbf{Definition 3.} The rate pair $(R_1,R_2)$ is said to be achievable for the CIC-PC if there exists a sequence of $(2^{nR_1},2^{nR_2},n)$ codes such that the average error probability $P_e^{(n)}=Pr(g_1(Y_1^n)\neq{(W_1)}~\mbox{or}~g_2(Y_2^n)\neq{W_2})\rightarrow{0}$ as n goes to infinity. The capacity is defined as the closure of the set of achievable rate pairs $(R_1,R_2)$.

\section{New Upper Bound for the CIC-PC}
In this section, we derive a new upper bound on the capacity region of the general CIC-PC. This new upper bound will be used later when we study the discrete memoryless semideterministic CIC-PC.

\vspace{1.5mm}
\textbf{Theorem 1.} \emph{Achievable rate pairs $(R_1,R_2)$ belong to the union of rate regions given by}
\begin{subequations}
\begin{align}
R_1&<I(X_1;Y_1|X_2,X_{r_1})\\
R_2&<I(V,X_2,X_{r_1};Y_2)\\
R_2&<I(X_1,X_2,X_{r_1};Y_2)\\
R_1+R_2&<I(V,X_2,X_{r_1};Y_2)+I(X_1;Y_1|V,X_2,X_{r_1})\\
R_1+R_2&<I(T,X_{1},X_{2};Y_{1}|X_{r_1})+I(V;Y_{2}|T,X_{r_1})\nonumber\\&\hspace{11mm}-I(V;Y_{1}|T,X_{r_1})\\
R_1+R_2&<I(X_1,X_2;Y_1,Y_2|X_{r_1})
\end{align}
\end{subequations}
\emph{where the union is over all joint probability mass functions $p(v,t,x_1,x_2,x_{r_1},y_1,y_2)$ which satisfies the Markov chain}
\begin{align}
(V,T)\rightarrow{(X_1,X_2,X_{r_1},Y_1)}\rightarrow{Y_2}.
\end{align}

\emph{Proof.} Consider a $(2^{nR_1},2^{nR_2},n)$ code with average error probability $P_e^{(n)}$. The probability distribution on the joint ensemble space $\mathcal{W}_1\times{\mathcal{W}_2}\times{\mathcal{X}_1^n}\times{\mathcal{X}_2^n}\times{\mathcal{X}}_{r_1}^n\times{\mathcal{Y}_1^n}\times{\mathcal{Y}_2^n}$ is given by
\begin{align}
&p(\omega_1,\omega_2,x_1^n,x_2^n,x_{r_1}^n,y_1^n,y_2^n)=\nonumber\\
&\hspace{11mm}p(\omega_1)p(\omega_2)p(x_1^n|\omega_1,\omega_2)p(x_2^n|\omega_2)\nonumber\\
&\hspace{11mm}\times\prod_{i=1}^n{p(x_{r_1,i}|y_1^{i-1})p(y_{1,i},y_{2,i}|x_{1,i},x_{2,i},x_{r_1,i})}.
\end{align}

By Fano's inequality, we have
\begin{subequations}
\begin{align}
H(W_i|Y_i^n)&\leq{nR_iP_e^{(n)}+1\triangleq n\delta_{i,n}},\mbox{~for~}i=1,2
%H(W_2|Y_2^n)&\leq{nR_2P_e^{(n)}+1\triangleq n\delta_{2,n}},
\end{align}
\end{subequations}
where $\delta_{1,n}$ and $\delta_{2,n}$ tends to zero when $n$ goes to infinity. We define the auxiliary random variables
\begin{align}
T_i&=(Y_1^{i-1},Y_{2,i+1}^n),\\
V_i&=(W_2,Y_1^{i-1},Y_{2,i+1}^n)
\end{align}

for $i\in\{1,...,n\}$. We first bound $R_1$ as follows:

\begin{align}
nR_1-&n\delta_{1,n}=H(W_1)-n\delta_{1,n}\leq{I(W_1;Y_1^n|W_2)}\\
&=\sum_{i=1}^n{I(W_1;Y_{1,i}|W_2,Y_1^{i-1})}\\
&=\sum_{i=1}^n{I(W_1;Y_{1,i}|W_2,Y_1^{i-1},X_{2,i},X_{r_1,i})}\\
&\leq{\sum_{i=1}^n{[H(Y_{1,i}|X_{2,i},X_{r_1,i})}}\nonumber\\
&\hspace{6mm}-H(Y_{1,i}|W_1,W_2,Y_1^{i-1},X_{1,i},X_{2,i},X_{r_1,i})]\\
&=\sum_{i=1}^n[H(Y_{1,i}|X_{2,i},X_{r_1,i})\nonumber\\
&\hspace{6mm}-H(Y_{1,i}|X_{1,i},X_{2,i},X_{r_1,i})]\\
&=\sum_{i=1}^n{I(X_{1,i};Y_{1,i}|X_{2,i},X_{r_1,i})},
\end{align}
where (7) follows since $W_1$ and $W_2$ are independent, (8) follows from the chain rule, (9) is due to the fact that $X_{r_1,i}$ is a deterministic function of $Y_1^{i-1}$ and $X_{2,i}$ is a deterministic function of $W_2$, (10) follows because conditioning does not increase the entropy, and finally (11) follows from the Markov chain $(W_1,W_2,Y_1^{i-1})\rightarrow(X_{1,i},X_{2,i},X_{r_1,i})\rightarrow{Y_{1,i}}$.
We then bound $R_2$ as follows:
\begin{align}
n&R_2-n\delta_{2,n}=H(W_2)-n\delta_{2,n}\nonumber\\
&\leq{I(W_2;Y_2^n)}=\sum_{i=1}^n{I(W_2;Y_{2,i}|Y_{2,i+1}^n)}\\
&\leq{\sum_{i=1}^n{H(Y_{2,i})-H(Y_{2,i}|Y_1^{i-1},Y_{2,i+1}^n,W_2)}}\\
&=\sum_{i=1}^n[H(Y_{2,i})\nonumber\\
&\hspace{11mm}-H(Y_{2,i}|Y_1^{i-1},Y_{2,i+1}^n,W_2,X_{2,i},X_{r_1,i})]\\
&=\sum_{i=1}^n{I(V_i,X_{2,i},X_{r_1,i};Y_{2,i})},
\end{align}
where (14) follows from the fact that conditioning does not increase the entropy, (15) follows because $X_{r_1,i}$ is a deterministic function of $Y_1^{i-1}$, and $X_{2,i}$ is a deterministic function of $W_2$, and (16) follows from the definition of $V_i$, given in (6). To establish the second bound on $R_2$, we have
\begin{align}
n&R_2-n\delta_{2,n}=H(W_2)-n\delta_{2,n}\nonumber\\
&\leq{I(W_2;Y_2^n)}=\sum_{i=1}^n{I(W_2;Y_{2,i}|Y_{2}^{i-1})}\nonumber\\
&\leq{\sum_{i=1}^n{H(Y_{2,i})-H(Y_{2,i}|Y_1^{i-1},Y_{2}^{i-1},W_1,W_2)}}\\
&=\sum_{i=1}^n[H(Y_{2,i})\nonumber\\
&\hspace{4mm}-H(Y_{2,i}|Y_1^{i-1},Y_{2}^{i-1},X_{1,i},X_{2,i},X_{r_1,i},W_1,W_2)]\\
&={\sum_{i=1}^n{H(Y_{2,i})-H(Y_{2,i}|X_{1,i},X_{2,i},X_{r_1,i})}}\\
&=\sum_{i=1}^n{I(X_{1,i},X_{2,i},X_{r_1,i};Y_{2,i})},
\end{align}
where (17) follows from the fact that conditioning does not increase entropy, (18) follows because $X_{1,i}$, $X_{2,i}$ and $X_{r_1,i}$ are deterministic functions of $(W_1,W_2)$, $W_2$ and $Y_1^{i-1}$, respectively; finally (19) follows from the Markov chain $(W_1,W_2,Y_1^{i-1},Y_2^{i-1})\rightarrow(X_{1,i},X_{2,i},X_{r_1,i})\rightarrow{Y_{2,i}}$.

For the sum rate, we now obtain the following bound:
\begin{align}
\hspace{-1mm}&n(R_1+R_2)-n(\delta_{1,n}+\delta_{2,n})\nonumber\\
&\leq{I(W_2;Y_2^n)+I(W_1;Y_1^n|W_2)}\\
&=\sum_{i=1}^n[I(W_2;Y_{2,i}|Y_{2,i+1}^n)+I(W_1;Y_{1,i}|W_2,Y_1^{i-1})]\\
%&\leq\sum_{i=1}^n[I(W_2,Y_1^{i-1};Y_{2,i}|Y_{2,i+1}^n)\nonumber\\
%&\hspace{2mm}-I(Y_1^{i-1};Y_{2,i}|W_2,Y_{2,i+1}^n)\nonumber\\
%&\hspace{2mm}+I(W_1,Y_{2,i+1}^n;Y_{1,i}|W_2,Y_1^{i-1})]\nonumber\\
&\leq\sum_{i=1}^n[I(W_2,Y_1^{i-1};Y_{2,i}|Y_{2,i+1}^n)-I(Y_1^{i-1};Y_{2,i}|W_2,Y_{2,i+1}^n)\nonumber\\
%&\hspace{2mm}-I(Y_1^{i-1};Y_{2,i}|W_2,Y_{2,i+1}^n)\nonumber\\
&\hspace{2mm}+I(Y_{2,i+1}^n;Y_{1,i}|W_2,Y_1^{i-1})\nonumber\\
&\hspace{2mm}+I(W_1;Y_{1,i}|W_2,Y_1^{i-1},Y_{2,i+1}^n)]\\
&=\sum_{i=1}^n[I(W_2,Y_1^{i-1};Y_{2,i}|Y_{2,i+1}^n)\nonumber\\
&\hspace{2mm}+I(W_1;Y_{1,i}|W_2,Y_1^{i-1},Y_{2,i+1}^n)]\\
&\leq\sum_{i=1}^n[I(W_2,Y_1^{i-1},Y_{2,i+1}^n,X_{2,i},X_{r_1,i};Y_{2,i})\nonumber\\
&\hspace{2mm}+I(W_1;Y_{1,i}|W_2,Y_1^{i-1},Y_{2,i+1}^n)]\\
&=\sum_{i=1}^n[I(W_2,Y_1^{i-1},Y_{2,i+1}^n,X_{2,i},X_{r_1,i};Y_{2,i})\nonumber\\
&\hspace{2mm}+I(W_1;Y_{1,i}|W_2,Y_1^{i-1},Y_{2,i+1}^n,X_{2,i},X_{r_1,i})]\\
&\leq\sum_{i=1}^n[I(W_2,Y_1^{i-1},Y_{2,i+1}^n,X_{2,i},X_{r_1,i};Y_{2,i})\nonumber\\
&\hspace{2mm}+H(Y_{1,i}|W_2,Y_1^{i-1},Y_{2,i+1}^n,X_{2,i},X_{r_1,i})\nonumber\\
&\hspace{2mm}-H(Y_{1,i}|W_1,W_2,Y_1^{i-1},Y_{2,i+1}^n,X_{1,i},X_{2,i},X_{r_1,i})]\nonumber
\end{align}
\begin{align}
&=\sum_{i=1}^n[I(W_2,Y_1^{i-1},Y_{2,i+1}^n,X_{2,i},X_{r_1,i};Y_{2,i})\nonumber\\
&\hspace{2mm}+H(Y_{1,i}|W_2,Y_1^{i-1},Y_{2,i+1}^n,X_{2,i},X_{r_1,i})\nonumber\\
&\hspace{2mm}-H(Y_{1,i}|W_2,Y_1^{i-1},Y_{2,i+1}^n,X_{1,i},X_{2,i},X_{r_1,i})]\\
&=\sum_{i=1}^n[I(V_i,X_{2,i},X_{r_1,i};Y_{2,i})\nonumber\\&\hspace{10mm}+I(X_{1,i};Y_{1,i}|V_i,X_{2,i},X_{r_1,i})],
\end{align}
where (21) follows as $W_1$ and $W_2$ are independent, (22) and (23) follow from the chain rule, (24) follows from applying Csiszar-Koener's Lemma \cite{c10} to (23); (25) is due to the fact that conditioning does not increase the entropy, (26) follows because $X_{r_1,i}$ is a deterministic function of  $Y_1^{i-1}$, and $X_{2,i}$ is a deterministic function of $W_2$, and finally (27) follows from the Markov chain $(W_1,W_2,Y_1^{i-1})\rightarrow(X_{1,i},X_{2,i},X_{r_1,i})\rightarrow{Y_{1,i}}$.

We now consider the second bound on the sum rate $R_1+R_2$:
\begin{align}\vspace{1mm}
n&(R_1+R_2)-n(\delta_{1,n}+\delta_{2,n})\nonumber\\
&=H(W_1)+H(W_2)-n(\delta_{1,n}+\delta_{2,n})\nonumber\\
&\leq{I(W_1;Y_1^n)+I(W_2;Y_2^n)}\\
&\leq{I(W_1;Y_1^n,W_2)+I(W_2;Y_2^n)}\nonumber\\
&={I(W_1;Y_1^n|W_2)+I(W_2;Y_2^n)+I(W_2;Y_1^n)-I(W_2;Y_1^n)}\nonumber\\
&={I(W_1,W_2;Y_1^n)+I(W_2;Y_2^n)-I(W_2;Y_1^n).}
\end{align}
For the first term in (30), we have:
\begin{align}
&I(W_1,W_2;Y_1^n)=\sum_{i=1}^n{I(W_1,W_2;Y_{1,i}|Y_{1}^{i-1},X_{r_1,i})}\\
&\leq\sum_{i=1}^n[H(Y_{1,i}|X_{r_1,i})\nonumber\\
&\hspace{11mm}-H(Y_{1,i}|Y_1^{i-1},X_{1,i},W_1,W_2,X_{2,i},X_{r_1,i})]\\
&=\sum_{i=1}^n[H(Y_{1,i}|X_{r_1,i})\nonumber\\
&\hspace{11mm}-H(Y_{1,i}|Y_1^{i-1},X_{1,i},X_{2,i},X_{r_1,i})]\\
&\leq\sum_{i=1}^n[H(Y_{1,i}|X_{r_1,i})\nonumber\\
&\hspace{11mm}-H(Y_{1,i}|Y_1^{i-1},Y_{2,i+1}^n,X_{1,i},X_{2,i},X_{r_1,i})]\\
&=\sum_{i=1}^n{I(T_i,X_{1,i},X_{2,i};Y_{1,i}|X_{r_1,i})}.
\end{align}
where (31) follows as $X_{r_1,i}$ is a deterministic function of  $Y_1^{i-1}$, (32) follows because conditioning does not increase the entropy, and (33) follows from the Markov chain $(W_1,W_2,Y_1^{i-1})\rightarrow(X_{1,i},X_{2,i},X_{r_1,i})\rightarrow{Y_{1,i}}$ ; finally (35) follows from (5). For the sum of the second and third terms in (30), we obtain the following bound:
\begin{align}
&I(W_2;Y_2^n)-I(W_2;Y_1^n)\nonumber\\&\hspace{2mm}=\sum_{i=1}^n[I(W_2;Y_{2,i}|Y_{2,i+1}^n)-I(W_2;Y_{1,i}|Y_1^{i-1})]
\end{align}
\begin{align}
&\hspace{0mm}=\sum_{i=1}^n[I(W_2,Y_1^{i-1};Y_{2,i}|Y_{2,i+1}^n)-I(W_2,Y_{2,i+1}^n;Y_{1,i}|Y_1^{i-1})\nonumber\\&\hspace{11mm}-I(Y_1^{i-1};Y_{2,i}|Y_{2,i+1}^n,W_2)\nonumber\\&\hspace{11mm}+I(Y_{2,i+1}^n;Y_{1,i}|Y_1^{i-1},W_2)]\\
&\hspace{0mm}=\sum_{i=1}^n[I(W_2,Y_1^{i-1};Y_{2,i}|Y_{2,i+1}^n)\nonumber\\&\hspace{11mm}-I(W_2,Y_{2,i+1}^n;Y_{1,i}|Y_1^{i-1})]\\
&\hspace{0mm}=\sum_{i=1}^n[I(W_2;Y_{2,i}|Y_1^{i-1},Y_{2,i+1}^n)-I(W_2;Y_{1,i}|Y_1^{i-1},Y_{2,i+1}^n)\nonumber\\&\hspace{11mm}+I(Y_1^{i-1};Y_{2,i}|Y_{2,i+1})-I(Y_{2,i+1}^n;Y_{1,i}|Y_1^{i-1})]\\
&\hspace{0mm}=\sum_{i=1}^n[I(W_2;Y_{2,i}|Y_1^{i-1},Y_{2,i+1}^n,X_{r_1,i})\nonumber\\&\hspace{11mm}-I(W_2;Y_{1,i}|Y_1^{i-1},Y_{2,i+1}^n,X_{r_1,i})]\\
&\hspace{0mm}=\sum_{i=1}^n[I(V_i;Y_{2,i}|T_i,X_{r_1,i})-I(V_i;Y_{1,i}|T_i,X_{r_1,i})]
\end{align}
where (36), (37) and (39) follow from the chain rule; (38) and (40) follow from applying Csiszar-Koener's Lemma \cite{c9} to (37) and (39), respectively, and finally (41) follows from the definitions of $T_i$ and $V_i$ given in (5) and (6), respectively.

Now by substituting (35) and (41) into (30), we obtain:
\begin{align}\vspace{1mm}
n&(R_1+R_2)\nonumber\\
&\leq\sum_{i=1}^n[I(T_i,X_{1,i},X_{2,i};Y_{1,i}|X_{r_1,i})+I(V_i;Y_{2,i}|T_i,X_{r_1,i})\nonumber\\&\hspace{11mm}-I(V_i;Y_{1,i}|T_i,X_{r_1,i})].
\end{align}
Finally, consider the following bound on the sum rate:
\begin{align}\vspace{1mm}
n&(R_1+R_2)-n\delta_{1,n}= H(W_1,W_2)-n\delta_{1,n}\nonumber\\
&\leq{I(W_1,W_2;Y_1^n,Y_2^n)}\nonumber\\
&=\sum_{i=1}^n{I(W_1,W_2;Y_{1,i},Y_{2,i}|Y_1^{i-1},Y_2^{i-1},X_{r_1,i})}\\
&\leq\sum_{i=1}^n[H(Y_{1,i},Y_{2,i}|X_{r_1,i})\nonumber\\
&\hspace{5mm}-H(Y_{1,i},Y_{2,i}|W_1,W_2,Y_1^{i-1},Y_2^{i-1},X_{1,i},X_{2,i},X_{r_1,i})]\nonumber\\
&=\sum_{i=1}^n{I(X_{1,i},X_{2,i};Y_{1,i},Y_{2,i}|X_{r_1,i})}.
\end{align}
where (43) follows from the fact that $X_{r_1,i}$ is a deterministic function of $Y_1^{i-1}$, and (44) follows from the Markov chain $(W_1,W_2,Y_1^{i-1},Y_2^{i-1})\rightarrow(X_{1,i},X_{2,i},X_{r_1,i})\rightarrow{Y_{1,i},Y_{2,i}}$.

Now, we define $T=(T_Q,Q)$, $V=(V_Q,Q)$, $X_{r_1}=X_{r_1,Q}$, $X_m=X_{m,Q}$ and $Y_m=Y_{m,Q}$ for $m=1,2$, where auxiliary random variable $Q$ is independent of every other random variable and is distributed uniformly over $\{1,2,...,n\}$. Following standard steps, it is straightforward to show that applying the defined random variables to (12), (16), (20), (28), (42) and (44) results in the single letter bounds presented in Theorem 1.

\section{Capacity Results for the CIC with PC Destinations}
In this section, we study two classes of the discrete memoryless CIC-PC. We first consider the degraded CIC-PC and characterize the capacity region of this channel. We then consider the semideterministic CIC-PC, and derive the capacity region for a variation of this channel which we call \emph{more capable} semideterministic CIC-PC. We use the following lower bound when we derive the capacity results in this section.

\vspace{1.5mm}
\textbf{Theorem 2.} \emph{The capacity region of the discrete memoryless CIC-PC contains the union of rate-pairs $(R_1,R_2)$ satisfying}
\begin{subequations}
\begin{align}
\hspace{-2mm}R_1&<I(X_1;Y_1|U,X_2,X_{r_1})\\
\hspace{-2mm}R_2&<\min\{I(U,V,X_2,X_{r_1};Y_2),I(U,V,X_2;Y_1|X_{r_1})\}\\
\hspace{-2mm}R_1&+R_2<{\mbox{RHS~of}~(45b)}+I(X_1;Y_1|U,V,X_2,X_{r_1}),
\end{align}
\end{subequations}
\emph{where the union is over all joint probability mass functions of the form}
\begin{align}
p(u,v,x_1,x_2,x_{r_1},y_1,y_2)&=p(x_{r_1})p(u,x_2|x_{r_1})p(v|u,x_2,x_{r_1})\nonumber\\
&\hspace{-20mm}\times p(x_1|u,v,x_2,x_{r_1})p(y_1,y_2|x_1,x_2,x_{r_1}).
\end{align}

\emph{Proof.} Due to space considerations, we present only the outline of the coding strategy. The codebook generation, encoding, and decoding steps are exactly similar to the steps followed in the proof of [6, Theorem 1]. The proof is based on the rate splitting [10] and superposition coding [11] at the cognitive source node, and decode and forward relaying scheme at the relay node. The primary source node independently encodes its message. We adopt the regular encoding/slide window decoding strategy for decode-and-forward relaying scheme [11].

\subsection{Degraded CIC with PC Destinations}
In this subsection, we characterize the capacity region of the discrete memoryless degraded CIC-PC.

\vspace{1.5mm}
\textbf{Definition 4.} A discrete memoryless CIC-PC is degraded if the channel transition probability distribution satisfies
\begin{align}
p(y_1,y_2|x_1,x_2,x_{r_1})=p(y_1|x_1,x_2,x_{r_1})p(y_2|y_1,x_{r_1}).
\end{align}

\vspace{1mm}
\textbf{Theorem 3.} \emph{For the discrete memoryless degraded CIC-PC, the capacity region is given by the union of rate regions}
\begin{subequations}
\begin{align}
\hspace{-2mm}R_1&<I(X_1;Y_1|U,X_2,X_{r_1})\\
\hspace{-2mm}R_2&<\min\{I(U,V,X_2,X_{r_1};Y_2),I(U,V,X_2;Y_1|X_{r_1})\}\\
\hspace{-2mm}R_1&+R_2<{\mbox{RHS~of}~(48b)}+I(X_1;Y_1|U,V,X_2,X_{r_1}),
\end{align}
\end{subequations}
\emph{where the union is over all joint probability mass functions of the form}
\begin{align}
p(u,v,x_1,x_2,x_{r_1},y_1,y_2)&=p(x_{r_1})p(u,x_2|x_{r_1})p(v|u,x_2,x_{r_1})\nonumber\\
&\hspace{-20mm}\times p(x_1|u,v,x_2,x_{r_1})p(y_1,y_2|x_1,x_2,x_{r_1}).
\end{align}

\emph{Proof.} Achievability follows from Theorem 2. For the converse proof see Appendix.

\textbf{Remark 1.} Theorem 3 reduces to the capacity region of the degraded partially cooperative relay broadcast channel given in [12] by setting $V=X_2=\emptyset$, $X_1=X$ and $X_{r_1}=X_1$.

\subsection{Semideterministic CIC with PC Destinations}
In this subsection, we characterize the capacity region of the more capable semideterministic CIC-PC.

\vspace{1mm}
\textbf{Definition 5.} A discrete memoryless CIC-PC is more capable if the channel satisfies
\begin{align}
I(V,X_2;Y_1|X_{r_1})>I(V,X_2,X_{r_1};Y_2)
\end{align}
for all $P_{VX_1X_2X_{r_1}Y_1Y_2}$ satisfying $V\rightarrow(X_1,X_2,X_{r_1})\rightarrow(Y_1,Y_2)$.

\vspace{1.5mm}
\textbf{Definition 6.} A discrete memoryless CIC-PC is semideterministic if the transition probability distribution $p(y_1|x_1,x_2,x_{r_1})$ takes on the values $0$ and $1$ only.

\vspace{1.5mm}
\textbf{Theorem 4.} \emph{For the discrete memoryless semideterministic CIC-PC satisfying the more capability condition given in Definition 5, the capacity region is given by the union of rate regions}
\begin{subequations}
\begin{align}
\hspace{-2mm}R_1&<H(Y_1|X_2,X_{r_1})\\
\hspace{-2mm}R_2&<I(V,X_2,X_{r_1};Y_2)\\
\hspace{-2mm}R_1&+R_2<I(V,X_2,X_{r_1};Y_2)+H(Y_1|V,X_2,X_{r_1}),
\end{align}
\end{subequations}
\emph{where the union is over all joint probability mass functions of the form}
\begin{align}
p(v,x_1,x_2,x_{r_1},y_1,y_2)&=p(x_{r_1})p(x_2|x_{r_1})p(v|x_2,x_{r_1})\nonumber\\
&\hspace{-20mm}\times p(x_1|v,x_2,x_{r_1})p(y_1,y_2|x_1,x_2,x_{r_1}).
\end{align}
%We first introduce the following lemma that will be useful for achievability proof.\vspace{1.5mm}

\emph{Proof.} Achievability follows from Theorem 2 by setting $U=\emptyset$, and then using
\begin{align}
&\min\{I(V,X_2,X_{r_1};Y_2),I(V,X_2;Y_1|X_{r_1})\}\nonumber\\&\hspace{18mm}=I(V,X_2,X_{r_1};Y_2),
\end{align}
which follows from satisfying more capability condition, and using $H(Y_1|X_1,X_2,X_{r_1})=0$ which follows from being semideterministic.
To prove the converse, we first prove that when the CIC-PC is semideterministic, the derived upper bound in Theorem 1 satisfies Markov chain $(V,T)\rightarrow{(X_1,X_2,X_{r_1})}\rightarrow{(Y_1,Y_2)}$:
\begin{align}
H(Y_2|V,T,X_1,X_2,X_{r_1})&=H(Y_2|V,T,X_1,X_2,X_{r_1},Y_1)\\
&=H(Y_2|X_1,X_2,X_{r_1},Y_1)\\
&=H(Y_2|X_1,X_2,X_{r_1}),
\end{align}
where (54) and (56) follows because $Y_1$ is a deterministic function of $(X_1,X_2,X_{r_1})$, and (55) follows from (2). Hence, (56) implies the Markov chain $(V,T)\rightarrow(X_1,X_2,X_{r_1})\rightarrow(Y_1,Y_2)$, i.e., for the semideterministic CIC-PC, $p(v,x_1,x_2,x_{r_1},y_1,y_2)$ satisfies the same Markov chain for both the upper bound (Theorem 1) and the lower bound (Theorem 2). Therefore, the converse follows from Theorem 1 using $H(Y_1|X_1,X_2,X_{r_1})=0$.\vspace{1.5mm}

\textbf{Remark 2.} The capacity of the ``semideterministic CIC-PC in the high-gain-interference regime" was derived in [2]. The ``more capable semideterministic CIC-PC" for which we derived the capacity region in Theorem 4 includes this channel as a special case. To prove this claim, we show that \emph{more capable} condition is weaker than the \emph{high-gain-interference} condition. The CIC-PC is in high-gain-interference regime if the channel satisfy
\begin{align}
I(X_2;Y_1|X_{r_1})&>I(X_2,X_{r_1};Y_2)\\
I(V;Y_1|X_2,X_{r_1})&>I(V;Y_2|X_2,X_{r_1})
\end{align}
for all $P_{VX_1X_2X_{r_1}Y_1Y_2}$ satisfying $V\rightarrow(X_1,X_2,X_{r_1})\rightarrow(Y_1,Y_2)$. Summing (57) and (58) results in (50). Therefore, the more capability condition stated in Definition 5 is weaker than the high-gain-interference condition. Hence, The "more capable semideterministic CIC-PC" is more general than the "semideterministic CIC-PC in high-gain-interference regime".

\section{Conclusions}
We derived a new upper bound on the capacity region of the discrete memoryless cognitive interference channel (CIC) with partially cooperative (PC) destinations (CIC-PC). In addition, we characterized the capacity region for the degraded CIC-PC as well as more capable semideterministic CIC-PC.

\appendix
First, we obtain the following upper bound on $R_1$:

\begin{align}
nR_1-&n\delta_{1,n}=H(W_1)-n\delta_{1,n}\leq{I(W_1;Y_1^n|W_2)}\\
&=\sum_{i=1}^n{I(W_1;Y_{1,i}|W_2,Y_1^{i-1})}\\
&=\sum_{i=1}^n{I(W_1,X_{1,i};Y_{1,i}|W_2,Y_1^{i-1},X_{2,i},X_{r_1,i})}\\
&={\sum_{i=1}^n{[H(Y_{1,i}|W_2,Y_1^{i-1},X_{2,i},X_{r_1,i})}}\nonumber\\
&\hspace{6mm}-H(Y_{1,i}|W_2,Y_1^{i-1},X_{1,i},X_{2,i},X_{r_1,i})]\\
&=\sum_{i=1}^n{I(X_{1,i};Y_{1,i}|U_i,X_{2,i},X_{r_1,i})},
\end{align}
where (59) follows as $W_1$ and $W_2$ are independent, (60) follows from the chain rule, and (61) follows because $X_{1,i}$, $X_{2,i}$ and $X_{r_1,i}$ are deterministic functions of $(W_1,W_2)$, $W_2$ and $Y_1^{i-1}$, respectively; (62) follows from the Markov chain $(W_1,W_2,Y_1^{i-1})\rightarrow(X_{1,i},X_{2,i},X_{r_1,i})\rightarrow{Y_{1,i}}$, and finally, (63) follows from the following definition:
\begin{align}
U_i = (W_2,Y_1^{i-1}).
\end{align}

We now establish the following upper bound on $R_2$:
\begin{align}
n&R_2-n\delta_{2,n}=H(W_2)-n\delta_{2,n}\nonumber\\
&\leq{I(W_2;Y_2^n)}=\sum_{i=1}^n{I(W_2;Y_{2,i}|Y_2^{i-1})}\nonumber\\
&\leq{\sum_{i=1}^n{H(Y_{2,i})-H(Y_{2,i}|Y_1^{i-1},Y_2^{i-1},W_2)}}\nonumber
\end{align}
\begin{align}
&=\sum_{i=1}^n[H(Y_{2,i})\nonumber\\
&\hspace{11mm}-H(Y_{2,i}|Y_1^{i-1},Y_2^{i-1},W_2,X_{2,i},X_{r_1,i})]\\
&=\sum_{i=1}^n{I(U_i,V_i,X_{2,i},X_{r_1,i};Y_{2,i})},
\end{align}

where (65) follows because $X_{2,i}$ and $X_{r_1,i}$ are deterministic functions of $W_2$ and $Y_1^{i-1}$, respectively, and (66) follows from (64) and the following definition:
\begin{align}
V_i = (W_2,Y_1^{i-1},Y_2^{i-1}).
\end{align}

To establish the second upper bound on $R_2$, we have:
\begin{align}
n&R_2-n\delta_{2,n}=H(W_2)-n\delta_{2,n}\nonumber\\
&\leq{I(W_2;Y_2^n)}\leq{I(W_2;Y_1^n,Y_2^n)}\nonumber\\
&=\sum_{i=1}^n{I(W_2;Y_{1,i},Y_{2,i}|Y_{1}^{i-1},Y_{2}^{i-1},X_{r_1,i})}\\
&=\sum_{i=1}^n{I(W_2;Y_{1,i}|Y_{1}^{i-1},Y_{2}^{i-1},X_{r_1,i})}\nonumber\\&\hspace{11mm}+\sum_{i=1}^n{I(W_2;Y_{2,i}|Y_{1}^{i-1},Y_{2}^{i-1},Y_{1,i},X_{r_1,i})}\\
&=\sum_{i=1}^n{I(W_2;Y_{1,i}|Y_{1}^{i-1},Y_{2}^{i-1},X_{r_1,i})}\\
&\leq{\sum_{i=1}^n{H(Y_{1,i}|X_{r_1,i})-H(Y_{1,i}|Y_1^{i-1},Y_{2}^{i-1},W_2,X_{r_1,i},X_{2,i})}}\nonumber\\
&=\sum_{i=1}^n{I(U_i,V_i,X_{2,i};Y_{1,i}|X_{r_1,i})},
\end{align}
where (68) follows because $X_{r_1,i}$ is a deterministic function of $Y_1^{i-1}$, (69) follows from the chain rule, and (70) follows from (47) which implies the Markov chain $(W_1,W_2,Y_1^{i-1},Y_2^{i-1})\rightarrow(X_{r_1,i},Y_{1,i})\rightarrow{Y_{2,i}}$.

We now derive the following upper bound on the sum rate:
\begin{align}
&n(R_1+R_2)-n(\delta_{1,n}+\delta_{2,n})\nonumber\\
&\leq{I(W_2;Y_2^n)+I(W_1;Y_1^n,Y_2^n|W_2)}\\
&=\sum_{i=1}^n[I(W_2;Y_{2,i}|Y_2^{i-1})\nonumber\\&\hspace{6mm}+I(W_1;Y_{1,i},Y_{2,i}|W_2,Y_1^{i-1},Y_2^{i-1},X_{2,i},X_{r_1,i})]\\
&=\sum_{i=1}^n[I(W_2;Y_{2,i}|Y_2^{i-1})\nonumber\\&\hspace{6mm}+I(W_1;Y_{1,i}|W_2,Y_1^{i-1},Y_2^{i-1},X_{2,i},X_{r_1,i})\nonumber\\&\hspace{6mm}+I(W_1;Y_{2,i}|W_2,Y_1^{i-1},Y_2^{i-1},Y_{1,i},X_{2,i},X_{r_1,i})]\\
&=\sum_{i=1}^n[I(W_2;Y_{2,i}|Y_2^{i-1})\nonumber\\&\hspace{6mm}+I(W_1;Y_{1,i}|W_2,Y_1^{i-1},Y_2^{i-1},X_{2,i},X_{r_1,i})]
\end{align}
\begin{align}
&\leq\sum_{i=1}^n[I(W_2,Y_1^{i-1},Y_2^{i-1},X_{2,i},X_{r_1,i};Y_{2,i})\nonumber\\&\hspace{6mm}+H(Y_{1,i}|W_2,Y_1^{i-1},Y_2^{i-1},X_{2,i},X_{r_1,i})\nonumber\\&\hspace{6mm}-H(Y_{1,i}|W_1,W_2,Y_1^{i-1},Y_2^{i-1},X_{1,i},X_{2,i},X_{r_1,i})]\\
&\leq\sum_{i=1}^n[I(W_2,Y_1^{i-1},Y_2^{i-1},X_{2,i},X_{r_1,i};Y_{2,i})\nonumber\\&\hspace{6mm}+H(Y_{1,i}|W_2,Y_1^{i-1},Y_2^{i-1},X_{2,i},X_{r_1,i})\nonumber\\&\hspace{6mm}-H(Y_{1,i}|W_2,Y_1^{i-1},Y_2^{i-1},X_{1,i},X_{2,i},X_{r_1,i})]\\
&=\sum_{i=1}^n[I(U_i,V_i,X_{2,i},X_{r_1,i};Y_{2,i})\nonumber\\&\hspace{6mm}+I(X_{1,i};Y_{1,i}|U_i,V_i,X_{2,i},X_{r_1,i})],
\end{align}
where (72) follows since $W_1$ and $W_2$ are independent, (73) follows because $X_{2,i}$ and $X_{r_1,i}$ are deterministic functions of $W_2$ and $Y_1^{i-1}$, respectively, and (74) follows from the chain rule; (75) follows from (47), (76) follows because conditioning does not increase the entropy, and finally (77) follows from the Markov chain $(W_1,W_2,Y_1^{i-1},Y_2^{i-1})\rightarrow(X_{1,i},X_{2,i},X_{r_1,i})\rightarrow{Y_{1,i}}$.

Finally, from (44) we have:
\begin{align}\vspace{1mm}
n&(R_1+R_2)-n\delta_{1,n}\leq\sum_{i=1}^n{I(X_{1,i},X_{2,i};Y_{1,i},Y_{2,i}|X_{r_1,i})}.
\end{align}


\begin{thebibliography}{1}
\bibitem{c1}
  N. Devroye, P. Mitran, and V. Tarokh, ``Achievable rates in cognitive channels,'' \emph{IEEE Trans. Info. Theory}, vol. 52, no. 5, pp. 1813-1827, May 2006.

\bibitem{c2}
  H. Y. Chu, and H. J. Su, ``On the capacity region of the cognitive interference channel with unidirectional destination cooperation,'' in \emph{Proc. IEEE Int. Symp. Inf. Theory}, pages 2408-2412, Jul. 2011.

\bibitem{c3}
  E. C. van der Meulen, ``Three-terminal communication channels,'' \emph{Advances in Applied Probability}, vol. 3, pp. 120-154, 1971.

\bibitem{c4}
  T. M. Cover and A. A. E. Gamal, ``Capacity theorems for the relay channel,'' \emph{IEEE Trans. Info. Theory}, vol. 25, no. 5, pp. 572-584, Sep. 1979.

\bibitem{c5}
  A. A. El Gamal and M. Aref, ``The capacity of the semideterministic relay channel,'' \emph{IEEE Trans. Info. Theory}, vol. 28,  no. 3, pp.536 1982.

\bibitem{c6}
  M. Kazemi, M. R. Aref, and M. A. Attari, ``Cooperative Relay Cognitive Interference Channel,'' \emph{In Proc. IEEE Inter. Conf. on Inf. Theory and Inf. Security}, pages 1064-1069, Sep. 2010.

\bibitem{c7}
  M. Kazemi, M. Mirmohseni, and M. R. Aref, ``Cooperative Relay Cognitive Interference Channels with Causal Channel State Information,'' In \emph{Proc. of IEEE Inter. Conf. on Inf. Theory and Inf. Security}, pages 1070-1075, Sep. 2010.

\bibitem{c8}
  M. Kazemi, A. Vosoughi, ``On the Capacity Region of the Partially Cooperative Relay Cognitive Interference Channel,'' in \emph{Proc. IEEE Int. Symp. Inf. Theory}, pages 2424-2427, Jul. 2013.

\bibitem{c9}
  I. Csiszar and J. Korner, ``Broadcast Channels with Confidential Messages,'' \emph{IEEE Trans. Info. Theory}, vol. 24, no. 3, pp. 339-348, May 1978.

\bibitem{c10}
  T. S. Han and K. Kobayashi, ``A new achievable rate region for the interference channel,'' \emph{IEEE Trans. Info. Theory}, vol. 27, no. 1, pp. 49–60, Jan. 1981.

\bibitem{c11}
  A. El Gamal and Y.-H. Kim, \emph{Network Information Theory}. Cambridge University Press, 2011.

\bibitem{c12}
  Y. Liang,V. V. Veeravalli, ``Cooperative Relay Broadcast Channels,'' \emph{IEEE Trans. Info. Theory}, vol. 53, no. 3, pp. 900–928, Mar. 2007.
\end{thebibliography}
\end{document}